\documentclass[journal]{vgtc}                

\usepackage{mathptmx}
\usepackage{graphicx}
\usepackage{times}

\onlineid{116}

\vgtccategory{Applications}

\vgtcinsertpkg

\preprinttext{Submitted to IEEE Visualization 2007 --- Applications track}

\title{Visualization of Cosmological Particle-Based Datasets}

\ifvgtcjournal

\author{Paul Arthur Navr\'{a}til \textit{Student Member, IEEE}, Jarrett L. Johnson, and Volker Bromm}
\authorfooter{
\item
  Paul Arthur Navr\'{a}til is with the Texas Advanced Computing Center at the University of Texas
  at Austin.  E-mail: pnav@tacc.utexas.edu.
\item
  Jarrett L. Johnson and Volker Bromm are with the Department of Astronomy at the University of
  Texas at Austin.  E-mail: [jljohnson, vbromm]@astro.as.utexas.edu
}

\else

\author{Paul Arthur Navr\'{a}til\thanks{e-mail: pnav@tacc.utexas.edu}\\ %
        \parbox{1.8in}{\scriptsize \centering Texas Advanced Computing Center \\ University of Texas at Austin}%
\and Jarrett L. Johnson\thanks{e-mail:jljohnson@astro.as.utexas.edu}\\ %
     \parbox{1.4in}{\scriptsize \centering Department of Astronomy \\ University of Texas at Austin} %
\and Volker Bromm\thanks{e-mail:vbromm@astro.as.utexas.edu}\\ %
     \parbox{1.4in}{\scriptsize \centering Department of Astronomy \\ University of Texas at Austin}}

\fi  

\ifvgtcjournal

\shortauthortitle{Navr\'{a}til \MakeLowercase{\textit{et al.}}: Visualization of Cosmological Particle-Based Datasets}

\fi 

\abstract{We describe our visualization process for a particle-based
simulation of the formation of the first stars and their impact on
cosmic history.  The dataset consists of several hundred time-steps of
point simulation data, with each time-step containing approximately
two million point particles.  For each time-step, we interpolate the
point data onto a regular grid using a method taken from the radiance
estimate of photon mapping~\cite{Jensen:1998:ESO}.  We import the
resulting regular grid representation into ParaView~\cite{paraview},
with which we extract isosurfaces across multiple variables.  Our
images provide insights into the evolution of the early universe,
tracing the cosmic transition from an initially homogeneous state to
one of increasing complexity. Specifically, our visualizations capture
the build-up of regions of ionized gas around the first stars, their
evolution, and their complex interactions with the surrounding matter.
These observations will guide the upcoming {\it James Webb Space
Telescope}, the key astronomy mission of the next decade.  }

\keywords{Interpolation, Isosurface, Astronomy, Cosmology.}

\CCScatlist{ 
  \CCScat{I.3.6}{Computing Methodologies}%
{Computer Graphics}{Methodology and Techniques};
  \CCScat{J.2}{Computer Applications}{Physical Sciences and Engineering}{Astronomy}
}

\begin{document}



\firstsection{Introduction}

\maketitle

One of the most important open questions in modern cosmology is to
understand how the first stars in the universe, formed a few 100
million years (100~Myr) after the Big Bang, ended the so-called
``cosmic dark ages''~\cite{bromm:2004}.  The first stars transformed
the early universe from its simple, homogeneous initial state to one
of increasing complexity, thus setting the stage for the entire
subsequent history of structure and galaxy formation.  This crucial
transformation is concerned with two interrelated processes: the
re-ionization of the universe, and the enrichment with heavy chemical
elements. The universe before the formation of the first stars
consisted of neutral, almost pure hydrogen and helium gas. Prior to
the first stars, there were no sources of ultraviolet (UV) photons
yet, which are required to ionize hydrogen, and the cosmic gas was
completely neutral at this early time. Observations, however, have
shown that the universe was again highly ionized, at least beginning
one billion years after the Big Bang. How this so-called re-ionization
happened is a primary focus of current cosmological debate. The
universe underwent a second formative transformation at the end of the
dark ages, when the pristine cosmic gas, containing only the hydrogen
and helium produced in the Big Bang, was enriched with heavy chemical
elements that were synthesized in the first stars, and subsequently
dispersed in extremely energetic supernova explosions. The first
stars, therefore, began the long nucleosynthetic process that resulted
in all the heavy elements that we find in our Solar System today.

Both of these cosmic transformations were highly complex, involving
the three-dimensional evolution of dark matter and gas, coupled
together by gravity.  Furthermore, studying the impact of the first
stars on their surroundings requires a radiation-hydrodynamics
calculation, which is at the frontier of what is currently feasible.
Progress can therefore only be made with sophisticated, large-scale
numerical simulations.  Specifically, we use a Lagrangian,
particle-based technique to evolve the cosmic gas, the so-called
smoothed particle hydrodynamics (SPH) algorithm. Gravitational forces,
acting on both gas and dark matter, are solved with a hierarchical
tree method. Our code has been parallelized with the Message-Passing
Interface (MPI) library, and performs well on large Beowulf-type
systems.

From this simulation, we want to produce images of smooth isosurfaces
that represent the various gas structures present.  The particle data
from the simulation has unknown structure, which hinders direct
extraction of isosurfaces.  Instead, we use an interpolation similar
to the three-dimensional radiance estimation technique from
photon-mapping~\cite{Jensen:1998:ESO} to interpolate the particle data to
the vertices of a regular grid.  In our process, the user may define
the grid resolution and sampling range, which allows data to be
represented at various levels of detail.  With data on a regular grid,
we can use any of a variety of open-source software tools to extract
and view the isosurfaces.  For the images presented here we use
ParaView~\cite{paraview}, which provides a feature-rich set of tools
to create and enhance our images, including isosurface extraction and
smoothing.

A key aspect of current research on the first stars is related to the
build-up and growth of bubbles of either ionized gas or hot, heavy
element-enriched gas.  Our visualizations are crucial to elucidate the
evolution of the bubbles, their complex interaction with the
surrounding medium, and the time-dependent topology of multiple bubble
growth and overlap.  Other visualization techniques do not produce
images that clearly demonstrate these phenomena. Our numerical
simulations, coupled with the effective visualization techniques that
we describe here, will allow us to make predictions for the {\it James
Webb Space Telescope (JWST)}, planned for launch in $\sim 2013$.  The
{\it JWST}, the key astronomy mission in the next decade, is extremely
sensitive at near-infrared (NIR) wavelengths, where most of the light
from the first stars, emitted at the source in the UV, but
subsequently redshifted by the cosmic expansion into the NIR, is
expected to reside.

The remainder of the paper is organized as follows: in
Section~\ref{sec:cosmo_background} we present the cosmological
motivation for our visualization; in Section~\ref{sec:vis_proc} we
describe our visualization process, and in Section~\ref{sec:results}
we discuss the insights made possible by our visualizations.  We
present related work in Section~\ref{sec:related_work} and future work
in Section~\ref{sec:conclusion}.

\begin{figure}[htb]
  \centering
  \includegraphics[width=3in]{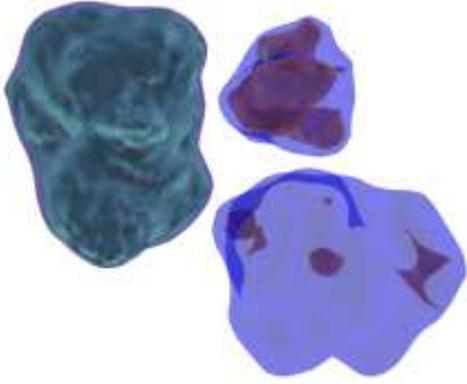}  
  \caption{Regions Ionized by the First Three Stars --- This figure
  shows the ionized gas within the first three ionized regions.  A
  region containing an active star is on the left, while in the two
  regions on the right the central stars have turned off and the
  electron fraction is dropping as the gas recombines.  Three
  isosurfaces of the electron fraction are shown: 0.34 (cyan), 0.10
  (orange) and 0.01 (blue).  The structures in the two right-most
  bubbles demonstrate that ionized gas cools and recombines
  non-uniformly, again becoming neutral.  It was often previously
  assumed that the gas recombined in a homogeneous fashion, with the
  electron fraction roughly equal throughout the ionized regions.  }
\label{fig:three_cores}
\end{figure}

\section{Cosmological Background}
\label{sec:cosmo_background}
\subsection{Simulations}

To study how the formation of the first stars affects the early
universe, we carry out detailed three-dimensional simulations of a
large volume of the cosmos during the epoch of the formation of the
first generation of stars.  These simulations allow us to follow the
evolution of the primordial gas as it collapses under the influence of
gravity to form stars, and as the radiation from these stars, in turn,
alters the chemistry and thermal state of the gas\cite{johnson:2007}.

\subsubsection{Smoothed Particle Hydrodynamics}

We carry out our simulations using the smoothed-particle hydrodynamics
(SPH) code GADGET, in which the gas is modeled as discrete particles,
each of which carries information about the dynamical, thermal, and
chemical properties of the gas at a given point~\cite{springel:2001,
springel:2002}. The mass contained in a given particle, however, is
smoothed out over a volume that depends on the mass density and the
numerical resolution. Variables at any given point in the fluid, and
not only at the location of the particles, can then be estimated by
kernel interpolation.  Our simulation uses $\sim$~2~million SPH
particles, along with $\sim$ 2 million dark matter particles, which
capture the gravitational effects that dark matter imposes.  The
volume of our cosmological simulation is a cubic box with side
length\footnote{The customary unit of distance in astronomy is the
parsec, where \mbox{1 pc $=$ $3.09\times 10^{16}$\,m $=$
3.26~lightyears}.} 460~kpc~$(1 + z)^{-1}$~$h^{-1}$, where $z$ is the
cosmological redshift, which decreases with time as the universe
expands, and $h = 0.7$ is the Hubble constant in units of
100~km~s$^{-1}$~Mpc$^{-1}$, which describes the rate of the cosmic
expansion.

We initialize our simulation at a redshift of $z = 99$, when the
primordial gas is nearly uniformly distributed across space and each
SPH particle modeling the gas has a density, temperature, and chemical
abundances as derived from Big Bang theory and cosmological
observations.  As the gas in our cosmological volume evolves, we track
the detailed evolution of the gas as it collapses, cools, and begins
to form stars at a redshift $z \sim 20$, or $\sim 200$~million~years
after the Big Bang.

\subsubsection{Radiation From the First Stars}

We model the effects of the radiation from the first generations of
stars by placing point sources of radiation at the sites in our
simulation where the gas collapses to high densities under the
influence of gravity.  The radiation from the first stars has two
important effects on their surroundings, the ionization and
concomitant heating of the primordial gas, and the destruction of the
H$_2$ molecules in the gas, which are important coolants allowing the
gas to collapse and form stars. The gas within a few kiloparsecs of
the star is ionized and heated to temperatures above $\sim 10^4$~K,
whereas the H$_2$ in the gas is destroyed within a slightly larger
region around the star.

We use a ray-tracing method to find the exact structure of these two
regions around the sites of star formation in our simulation. Within
the so-called H~II region, inside of which the gas is completely
ionized, we set the electron fraction of the gas to unity and raise
the temperature of the gas accordingly.  Within the Lyman-Werner (LW)
bubble, the region in which the molecules in the gas are destroyed, we
set the fraction of H$_2$ to zero.  Because the first stars are
expected to have been very massive, with masses perhaps 100~times that
of the Sun, we take it that the stars in our simulation have
correspondingly short lifetimes of 3~Myr.  Thus, after this brief
stellar lifetime has elapsed in our simulation, we allow the gas to
evolve once more without any radiative effects.  In our simulation,
eight stars are formed in the course of 50~Myr, and we carry out this
detailed process of including the radiative effects for each of these
stars.

\subsection{Data Types}

Each of the SPH particles that we use to model the primordial gas
carries all of the relevant information that we would like to know
about the dynamical, thermal, and chemical properties of the gas.  In
particular, each of these particles tracks the location, temperature,
density, molecule (H$_2$) fraction, and electron fraction of a given
parcel of gas.  It is these properties of the primordial gas which are
perhaps the most important to know in order to understand the process
of the formation of the first stars and protogalaxies.  Therefore, we
have focused on visualizing these properties of the gas as they vary
within our cosmological box, in order to extract an understanding of
some of the many phenomena that arise with the feedback imposed by the
first generations of stars.

\begin{figure*}[tb]
  \centering
  \includegraphics[width=2.25in]{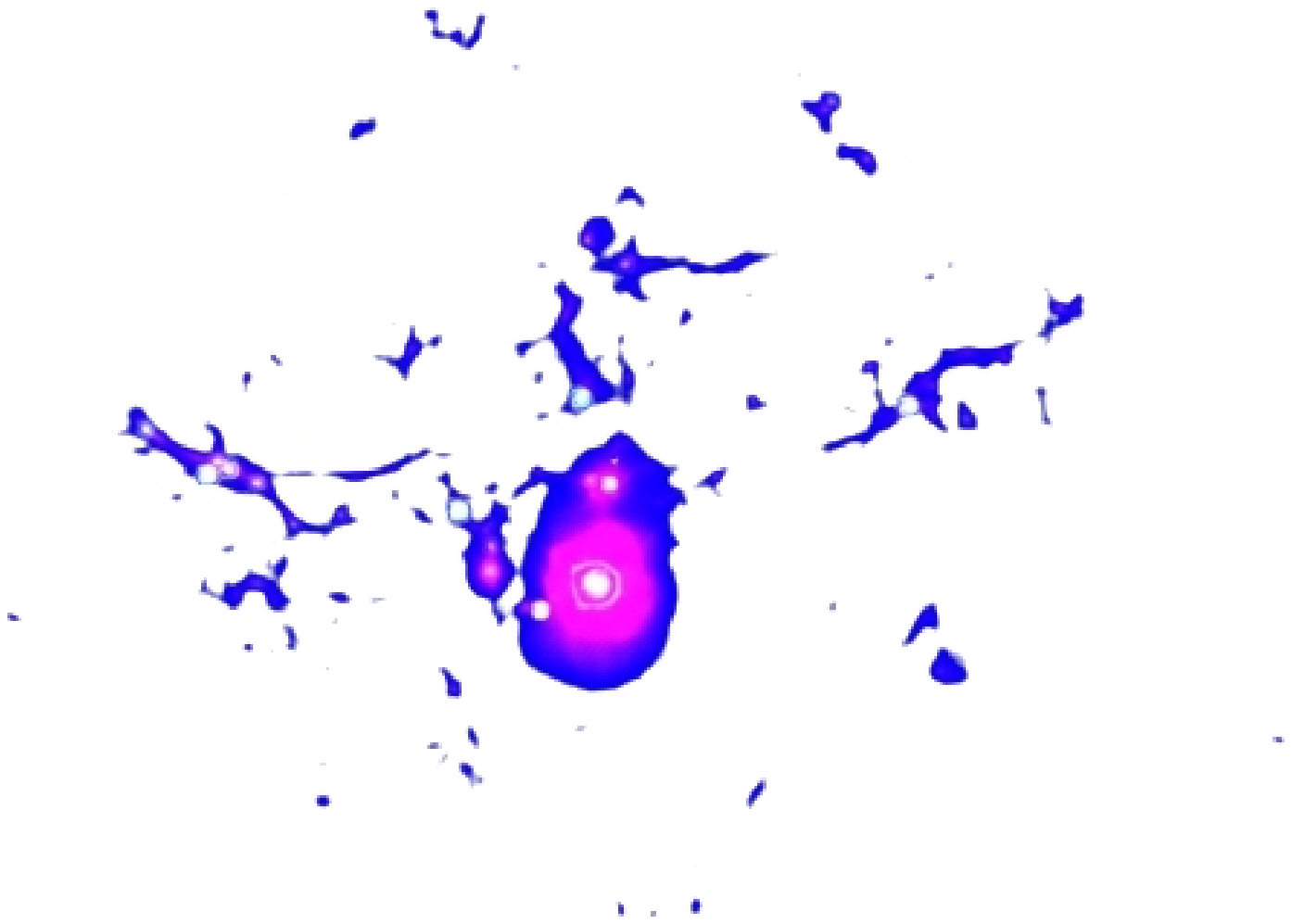}
  \includegraphics[width=2.25in]{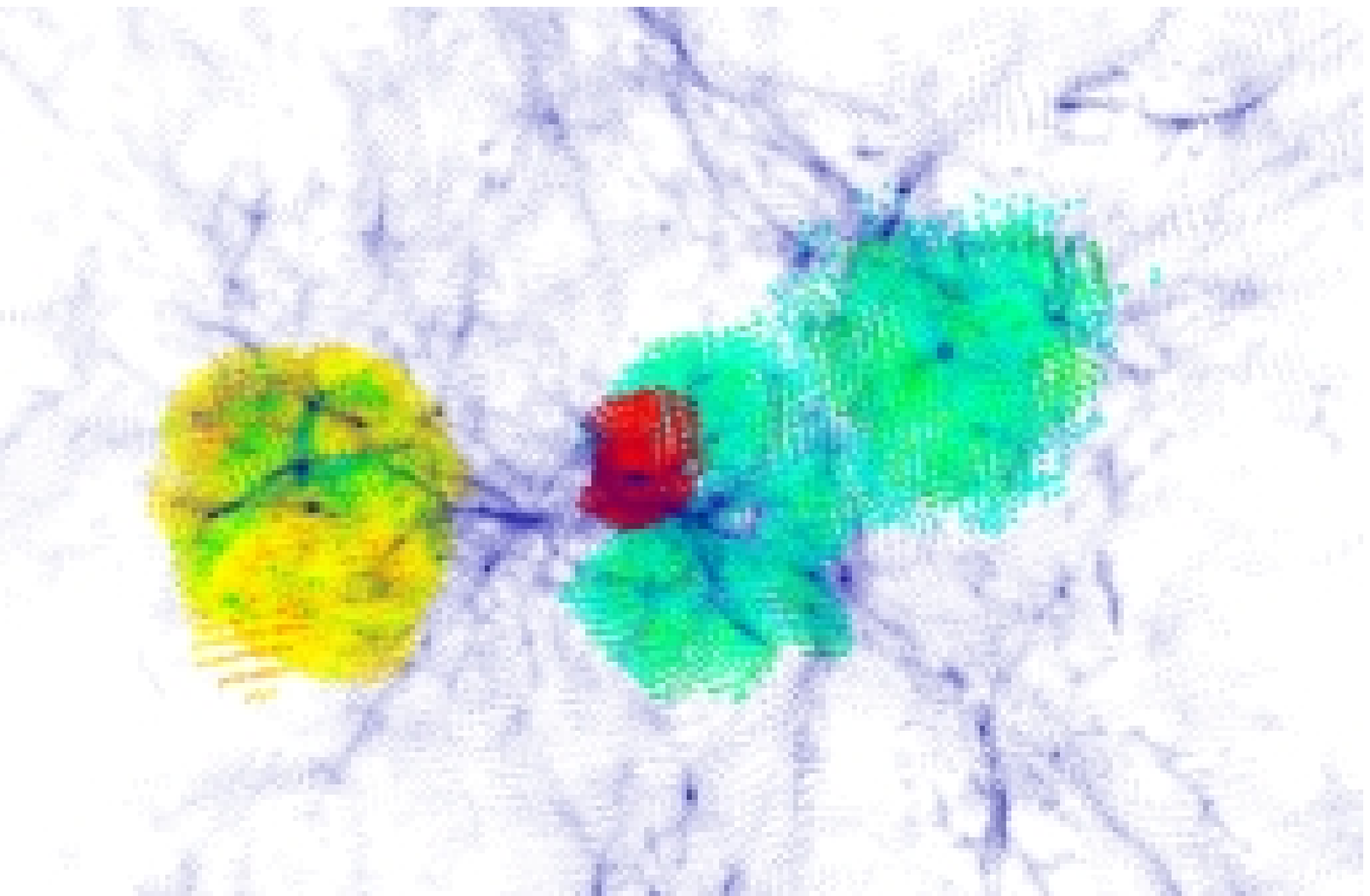}
  \includegraphics[width=2.25in]{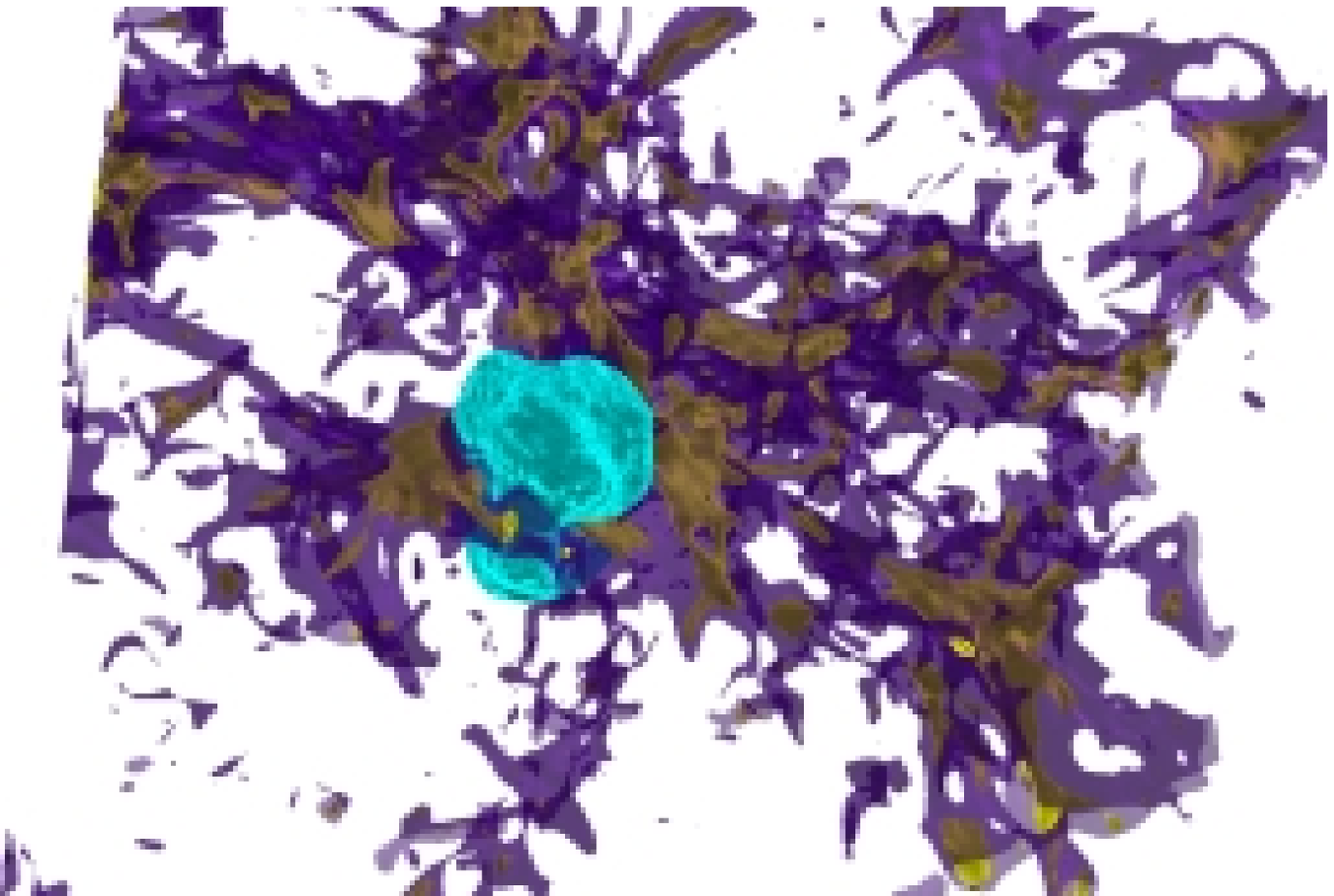}
  \caption{Technique Comparison --- here we compare the image quality
    between direct visualizations, which are often used for
    cosmological data sets, and our isosurface visualization.  All
    three images show the density field in our simulation, while the
    middle and right images also show the electron fraction in ionized
    regions.  The left image was generated directly from our particle
    data using alpha-blended sprites in Partiview~\cite{partiview}.
    The middle and right images were generated in
    ParaView~\cite{paraview}.  The middle image is a direct
    visualization of the particle data as points.  The right image was
    generated by resampling our particle data to a regular grid using
    the interpolation we describe in Section~\ref{sec:vis_proc}.  We
    then use ParaView~\cite{paraview} to extract and smooth the
    isosurfaces.  In the right image, the full extent of the gas
    structures are visible, and the spatial relationships among the
    structures are clearly defined.  Note that the transparency of the
    isosurfaces provides additional structural information without
    sacrificing the clarity of the spatial relationships.  {\it
    Partiview image courtesy of Karla Vega, Texas Advanced Computing
    Center}.}
\label{fig:technique_comp}
\end{figure*}

\section{Visualization Process}
\label{sec:vis_proc}

The most effective visualization of our particle simulation will
show the detailed structure of the gases and their spatial
relationships.  In particular, we would like to observe whether the
ionized ``bubbles'' form in regions of high molecular density, and if
hydrogen is suppressed in areas of active ionization.  We would also
benefit if the visualization revealed details of our simulation that
we did not expect.  Further, we seek a visualization process that
is quick to implement and utilizes as much existing freely available,
open-source software as possible.

In this section, we describe both our visualization techniques and the
considerations that motivated the techniques.  We hope that in so
doing we can inform the decision-making process of other researchers
faced with similar challenges.  We first present other visualization
techniques we considered and the reasons why we did not ultimately use
them.  We follow with the presentation of our chosen visualization
techniques.  We describe the cosmological insights made possible by
our visualizations in Section~\ref{sec:results}.

\subsection{Other Techniques Considered}

We considered several visualization techniques other than our
technique described in Section~\ref{sec:our_technique}.  We
discuss both {\em direct point visualization} techniques and 
other {\em isosurface extraction} techniques.

\subsubsection{Direct Point Visualization Techniques}

We tested two forms of direct point visualization: visualizing the
simulation particles as points using ParaView~\cite{paraview} and
visualizing the particles with alpha-blended sprites at each particle
location using Partiview~\cite{partiview}.
Figure~\ref{fig:technique_comp} contains example images from each
technique.  Because of the discrete nature of the particle simulation,
these visualizations only imply the extent of the gas structures we
wish to observe.  The visualizations lack the clear structural detail
necessary for scientific work.  It is also difficult to see the
spatial relationships among features, even when interacting with the
visualization in three dimensions.

To define the complete extent of the gas structures in our simulation,
we must select an isosurface extraction technique.

\subsubsection{Isosurface Extraction Techniques}
\label{sec:iso_extract_techniques}

Most isosurface extraction techniques operate only on a grid, whether
structured or unstructured~\cite{livnat:2005}.  Our particle
simulation produces a point cloud with no explicit order or
connectivity information, which complicates using an unstructured grid
representation for isosurface extraction.  Techniques
exist~\cite{co:2005} to extract isosurfaces directly from point
clouds, but we are unaware of any open-source implementations.  For
our purposes, the potential information gain from such techniques does
not overcome the implementation cost.  We can achieve sufficient
visualization quality from isosurface techniques that operate on a
regular grid.  Therefore, we will focus on resampling our data to a
regular grid.

While there are many isosurface extraction techniques, we would like
to use one with a well-tested, freely-available implementation.  If
such an implementation can meet our needs, we can eliminate potential
delays and errors from implementing a technique ourselves.  Ideally,
the implementation would be part of a complete visualization package
so we may immediately proceed with data exploration rather than
implementing image rendering and user interface code.

While many visualization packages implement isosurface extraction
techniques, few contain any resampling methods.  The Visualization
ToolKit (VTK)~\cite{vtk} contains only two, the most appropriate of
which is an implementation of Shepard's method~\cite{shepard:1968}.
However, this interpolation is a global technique, which means every
point in the original data set is included in the computation for each
interpolation point.  This process is order $n^2$, where $n$ is the
number of points in the original data set, and is unacceptably slow
when $n$ is large.  We want an interpolation that provides acceptable
fidelity to the original data along with the ability to control how
much of the original data is included in the computation for each 
interpolation point.

\begin{figure}[htb]
  \centering
  \includegraphics[width=2.5in]{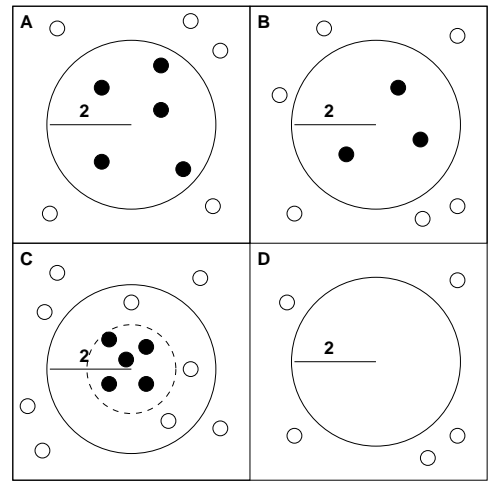}
  \caption{Our Local Interpolation Method --- we show four toy cases
    to illustrate how our interpolation method works for distance $d =
    2$ and $n = 5$, projected into two dimensions.  In ({\bf A}),
    exactly five particles are within the sampling radius $d$, and
    these five particles are used in the interpolation.  In ({\bf B}),
    only three particles are within the sampling radius, so only these
    three particles are used.  In ({\bf C}), there are eight particles
    within the sampling radius, so only the five ($n=5$) closest are
    used.  In ({\bf D}), no particles are within the sampling radius,
    so a set of default values are put at the interpolation point.
    For the results presented here, we used zeros for the scalar variables
    at such points.}
\label{fig:photon_map}
\end{figure}

\subsection{Our Visualization Technique}
\label{sec:our_technique}

Our chosen visualization technique borrows an interpolation used in
computer graphics to resample our particle data to a regular grid, 
then uses a well-tested, freely available visualization package
to extract isosurfaces and create visualization images.  While a 
quantitative analysis of our technique is beyond the scope of this 
paper, we provide the rationale that motivates our parameter 
choices and the impact different values would have.  

Our interpolation comes from the radiance estimation technique in
Jensen's photon mapping algorithm~\cite{Jensen:1996:GIU}, a global
illumination technique in computer graphics.  In photon mapping,
radiance samples are propagated from light sources to surfaces in a
scene prior to any visibility calculations.  The radiance at each
sample is calculated using the luminance properties of the light
source, the surface material properties, and the distance and angle
between light and surface.  These radiance samples are stored in a
spatial acceleration structure.  When rendering the scene, the
indirect diffuse illumination on a surface is estimated by
interpolating among the radiance samples near the visibility sample
point.  Jensen and Christensen~\cite{Jensen:1998:ESO} extend this
technique to sample radiance in three dimensions, both on surfaces and
in participating media.  We adapt this three-dimensional interpolation
technique to resample our particle data to a regular grid.

This sort of interpolation has been described as a localized inverse 
distance-weighted method~\cite{franke:1982} and as an $n^{th}$-nearest
neighbor density estimate~\cite{Jensen:1998:ESO}.

Our interpolation process performs the following steps: 1) determine
grid resolution; 2) insert particles into grid; and 3) interpolate
data to grid vertices.  We input the resulting regular grid
into our chosen visualization package to extract isosurfaces and
create the visualization images.

\subsubsection{Determining Grid Resolution} 

The grid resolution of the interpolation directly affects the
information quality of the visualization and the efficiency of the
visualization process.  An overly coarse grid will not capture
high-frequency details in the original data set, while an overly fine
grid will consume extra memory and computing resources without
expressing any new information.  The ideal grid resolution is no more
than half the size of the smallest expected
feature~\cite{Cook:1986:SSI}.

If we do not know the ideal grid resolution, we calculate a reasonable
grid resolution using the global point density in the original data
set.  First, we determine a reasonable number of grid cells per unit
distance ($cpud$) with the equation~\cite{pharr:2004}

\begin{equation}
  cpud = \frac{c\,n^{\frac{1}{3}}}{\Delta_{major\_axis}}\,length_{major\_axis}
\end{equation}

where $n$ is the number of particles in the original data set,
$\Delta_{major\_axis}$ is the length of the longest axis of a
bounding box around the original data set, and $c$ is an experimentally
derived constant.  For this paper, we found $c=1$ provides sufficient
grid resolution to capture relevant features in our data.  

To calculate the grid resolution, our process takes the bounding box
around the original data set and multiplies the length of each size by
$cpud$.  This makes the resulting cells as cube-like as possible so
that our interpolation is as uniform as possible.  

For initial data exploration, we can use an {\it artificially coarse}
grid resolution to create a smaller data set.  Coarse grids have less
data and thus isosurfaces can be extracted and visualized more
quickly.  Visualizing such a 'prototype' grid can be a useful way to
determine exact isovalues and view positions before visualizing for
all desired details on a finer grid.

For this paper, we used a $128 \times 128 \times 128$ grid, which
produced good visualizations in a reasonable amount of time.  Higher
resolutions increased the data size and processing time without a
significant increase in visualization quality, while lower resolutions
produced insufficiently accurate visualizations.  Note that all axes
contain the same number of cells because the extent of our data set
is cube-shaped.  If the extent of a data set is more rectangular, the grid
axes may have different number of cells.

\subsubsection{Particle Insertion}

Once the grid resolution has been determined, our process inserts
particles into grid cells according to their location in space.  The
insertion process is similar to inserting geometry in a grid
acceleration structure in computer graphics~\cite{Fujimoto:1985:ART}.
Since our interpolation points are located at the grid vertices, our
process uses the cells near a vertex to quickly eliminate particles
that need not be included in the interpolation for that vertex.  Our
process can eliminate from consideration all particles in a cell that
falls outside the local interpolation region (see
Section~\ref{sec:interpolation} below).

\subsubsection{Interpolation}
\label{sec:interpolation}

After the particles have been inserted into the grid, we perform an
inverse-distance weighted interpolation at each grid vertex.  For each
vertex, we would like to achieve a sufficiently-accurate interpolation
while using a minimal number of particles.  Specifically, we wish to
eliminate distant particles that only marginally affect the
interpolation result.  Thus we can calculate a close approximation of
a global interpolation with much less computation.

We control the number of particles used in the interpolation by
specifying both an inclusion distance $d$ for particles around the
interpolation point and a maximum number of particles $n$ that can be
included in the interpolation at each point.  The distance parameter
$d$ must be set large enough so that we expect a sufficient number of
particles to be included for the interpolation steps.  Yet, a
reasonable value for $d$ may include far more particles than expected
in a region of high particle density.  Therefore, we also set a
maximum number of particles $n$ that can be included in the interpolation,
so that the computation is still efficient.  For such cases, we include
the closest $n$ particles to the interpolation point.  Because our process
enforces the $n$ particle limit only in dense regions, we 
do not expect to compromise the quality of the interpolation.  The limit
only enforces the expected particle count implied by $d$.

Note that if a global interpolation is desired, $d$ can be set equal
to the major axis of the bounding box of the original data set and $n$
can be set equal to the total number of particles in the original data
set.  Thus, every particle would be included in the interpolation at
each interpolation point.

Figure~\ref{fig:photon_map} demonstrates the four possible
ways in which $d$ and $n$ control the number of particles included at
each interpolation point.  For the case where no particles fall 
within the inclusion distance $d$ (case {\bf D} in the figure), a 
default value for each interpolation variable is used.  For this paper,
we use zero.

Values of $d$ and $n$ should be selected according to the quality of
the interpolation desired.  Larger values for $d$ globally improve the
interpolation, and larger values for $n$ improve the interpolation for
dense regions of the original data set.  For this paper, we set $d$
equal to the width of one grid cell ($d = 3.62$) and we set $n$ equal
to one percent of the total particle count ($n \approx 20K$).  These
values are experimentally determined, where larger values do not
noticeably improve the quality of the interpolation for our purposes.

\subsubsection{Isosurface Extraction and Visualization}

After the interpolation is performed, we import the resulting regular
grid into ParaView~\cite{paraview} to extract and smooth the
isosurfaces.  We choose ParaView because it is a well-tested,
full-featured and open-source visualization package.  Thus we can
begin visualizing our data immediately rather than spending additional
effort implementing and debugging an ad-hoc solution.

\begin{figure}[htb]
  \centering
  \includegraphics[width=3in]{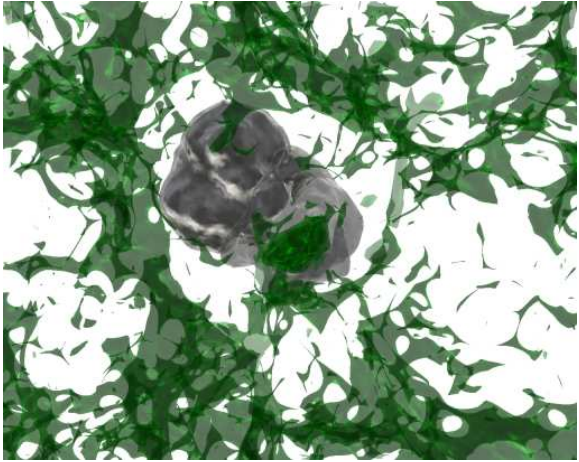}  
  \caption{Hydrogen Ionization --- This figure shows an active and a
    ``ghost'' ionized region, where recombination is taking place and
    the gas is again becoming neutral, along with the molecular
    hydrogen (H$_2$) fraction within our simulation.  Ionized gas
    isosurfaces $5e^{-2}$ (opaque) and $5e^{-3}$(transparent) are in
    gray, and hydrogen density isosurfaces $3e^{-6}$, $1e^{-5}$, and
    $1e^{-4}$ are in green Notice that the H$_2$ fraction is
    suppressed around the active ionized region, but is instead
    elevated within the ``ghost'' ionized region.}
\label{fig:h2frac}
\end{figure}

\begin{figure}[htb]
  \centering
  \includegraphics[width=3.1in]{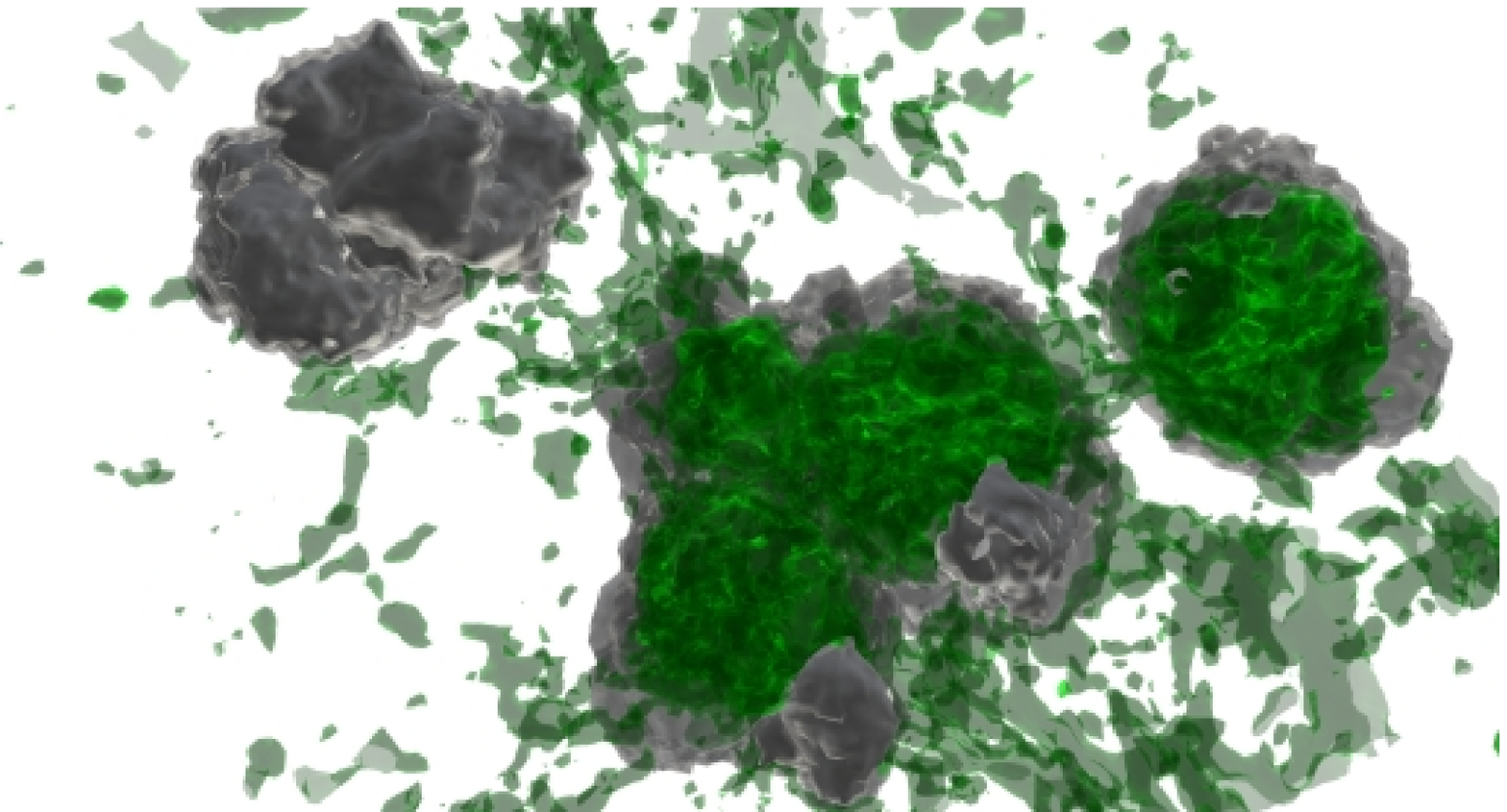}   
  \includegraphics[width=2in]{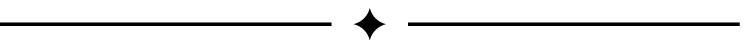}
  \includegraphics[width=3.1in]{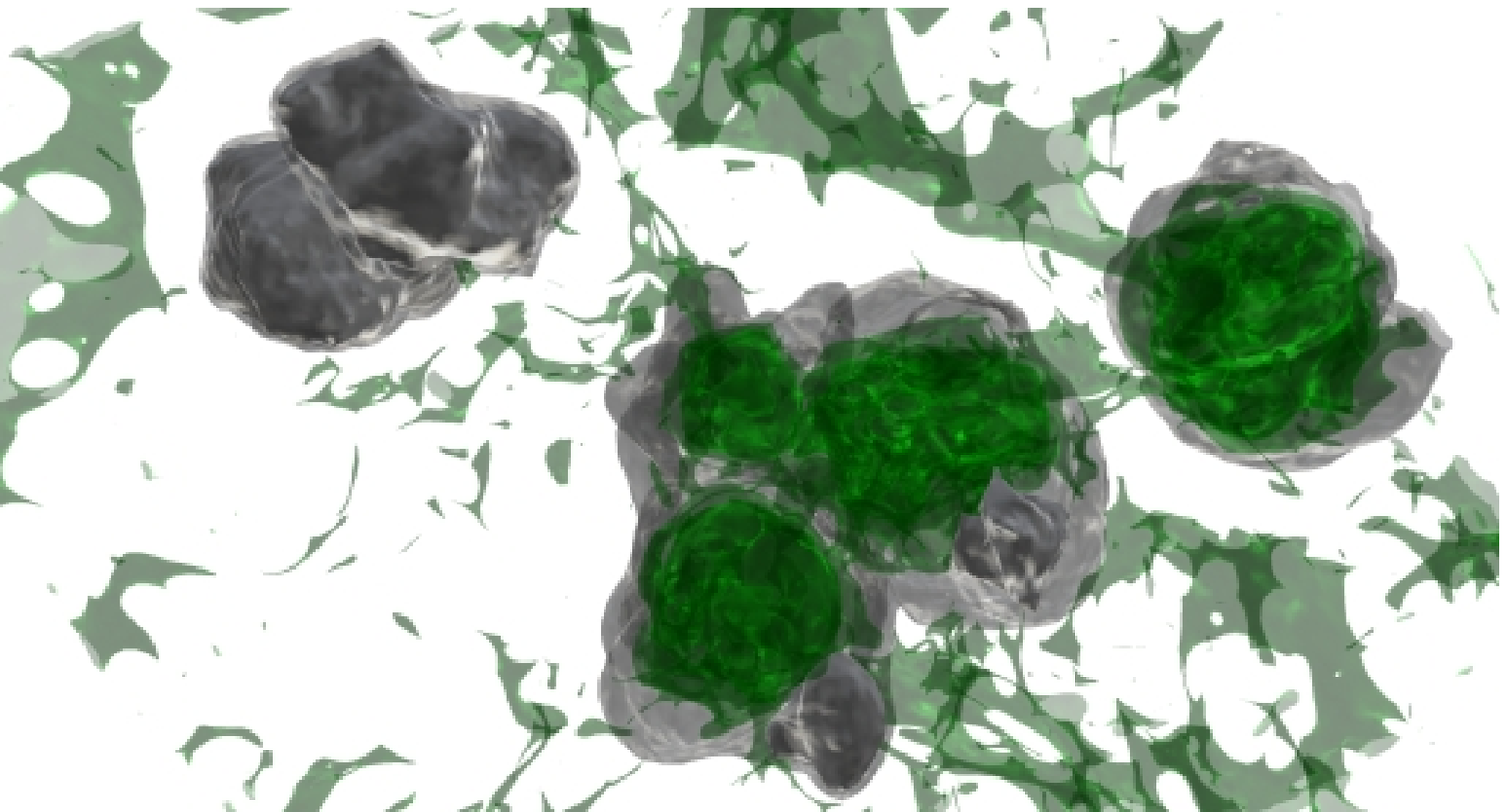}   
  \caption{Isosurface Extraction and Smoothing --- These figures show
    several active and one ``ghost'' ionized region, wherein
    recombination is taking place and the gas is again becoming
    neutral, along with the molecular hydrogen (H$_2$) fraction within
    our simulation.  The top image contains raw isosurfaces, and the
    bottom image contains isosurfaces after the application of a
    smoothing filter.  The smoothing filter sharpens the significant
    structures and removes the insignificant structures, making the
    spatial relationships among the structures more readily apparent.
    }
\label{fig:rough_smooth}
\end{figure}

\begin{figure}[htb]
  \centering
  \includegraphics[width=3.1in]{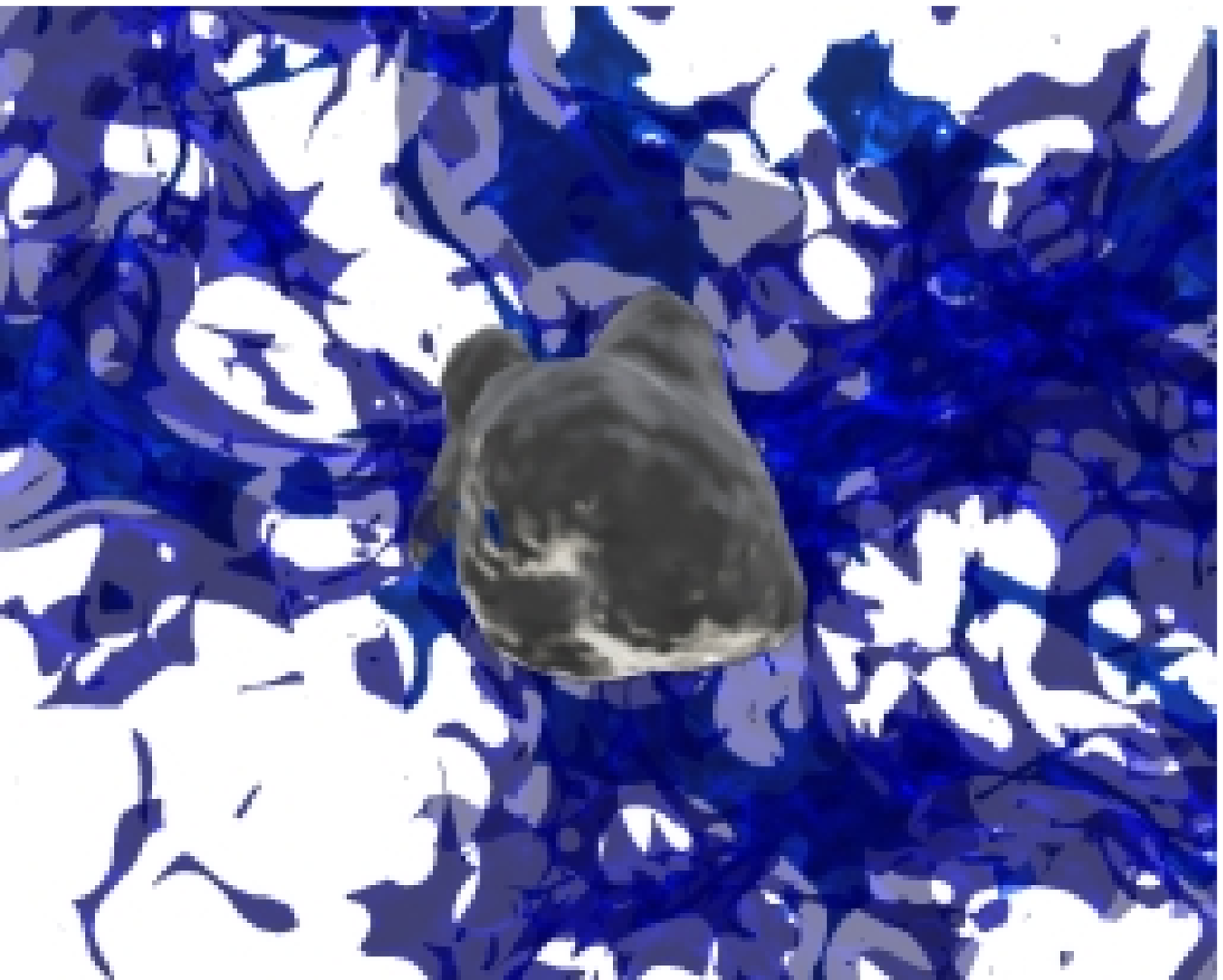}    
  \includegraphics[width=2in]{diamondrule.eps}
  \includegraphics[width=3.1in]{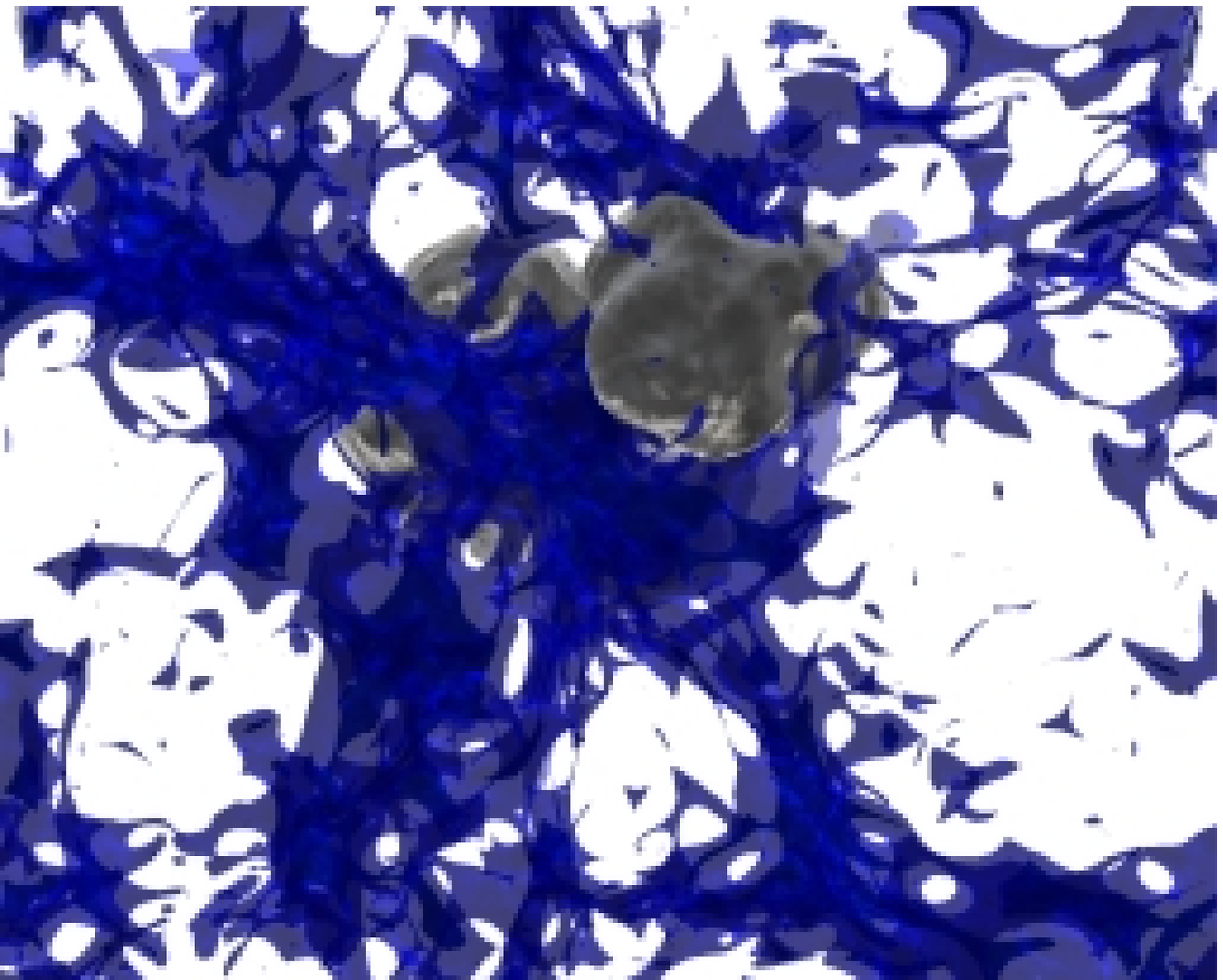}   
  \caption{Formation of Ionized Regions --- This figure shows the
    density field when the first star has formed (top) and after a
    second and third star have formed (bottom).  Notice that in the
    bottom image, the first star has turned off and only the ``ghost''
    ionized region is present (partially obscured by the molecular
    density isosurfaces).  Ionized gas isosurfaces $5e^{-2}$ (opaque)
    and $5e^{-3}$(transparent) are in gray, and density isosurfaces
    $1.1e^{-2}$ and $5e^{-3}$ are in blue.  Notice that the second and
    third stars have formed in the densest regions, as expected.  }
\label{fig:density}
\end{figure}

We use Marching Cubes~\cite{Lorensen:1987:MCA} to generate our
isosurfaces.  To ensure all isosurfaces are visible, we create only
two to three isosurfaces per visualized variable.  Low-value
isosurfaces are colored darker and have greater transparency (lower
alpha), whereas high-value isosurfaces are colored lighter and have
greater opacity (higher alpha).  For two-isosurface images, we use
alpha values of $0.4$ and $0.8$; for three-isosurface images, we use
alpha values of $0.3$, $0.5$, and $0.8$.  We select power-of-two
increments for isosurface hue (e.g., 64, 128, 255), to visually
distinguish isosurfaces of the same variable.  In general, we use a
standard color scheme to easily identify the variables in the
visualization (green for hydrogen density, blue for molecular density,
gray for ionized molecules; see
Figures~\ref{fig:h2frac},~\ref{fig:rough_smooth},~and~\ref{fig:density});
however, for dramatic effect in stand-alone images, we select
high-contrast hues (see
Figures~\ref{fig:three_cores}~and~\ref{fig:technique_comp}).

After creating the isosurfaces, we apply a smoothing filter to remove
noise and insignificant surfaces from the isosurface extraction.  We
use 1000 iterations of the smoothing filter for the images in this
paper.  In Figure~\ref{fig:rough_smooth}, we demonstrate the
improvement in image quality that the smoothing filter provides.  It
is easier to see the significant structures in the smoothed image, and
the spatial relationships among the structures are more readily
apparent.

ParaView also incorporates animation controls, with which we create
image sequences across our data sets.  By creating a time series in
ParaView, we create the desired isosurfaces for one time step and
generate tens to hundreds of images across the sequence.  The video
submitted with this paper was assembled from frames generated by
ParaView.

\section{Results and Discussion}
\label{sec:results}

The visualizations produced by our technique provide us with critical
insight into the structure and spatial organization of the gases in
our simulation, which furthers our cosmological understanding and the
impact of our work.  We describe several of these cosmological
insights below.  For more detail about the cosmology, we refer readers
to Johnson et al.~\cite{johnson:2007}.


\subsection{Cosmological Insights}

The key question for cosmology is to understand the fundamental
transition in the early universe from simplicity to complexity,
brought about by the first stars. Specifically, how was the universe
re-ionized, and how was it enriched with the first heavy elements?
Understanding the build-up of ionized regions around the first stars
is important for a number of reasons. Firstly, the evolution of the
ionized bubbles, the speed with which they grow, and the final size
reached, are diagnostic of the properties of the first stars, in
particular their mass. The typical mass of the first stars is
currently only predicted by theory, and it is essential to devise
empirical tests. One such test is to carry out simulations of the
early re-ionization process, assuming different stellar masses, and to
trace and characterize the resulting pattern of ionized bubbles.  The
observational signature includes the three-dimensional arrangement of
the bubbles, their clustering properties, and their overall volume
filling fraction as a function of time. Visualizations are crucial to
glean this signature from the numerical simulations.  Our
visualizations have shown that the bubble interior is not uniformly
ionized (see Figure~\ref{fig:three_cores}); instead there are pockets
of high-density gas inside of them that remain substantially
neutral. We find that higher mass stars tend to result in more
homogeneous bubble interiors. In addition, higher mass stars produce
larger ionized bubbles that fill a larger fraction of the available
cosmic volume.  Note that the structural detail found in our
isosurface visualizations enables these observations, observations
that are not possible with direct visualization on the particles.

Secondly, the build-up of bubbles is related to what is termed
``radiative feedback''. The basic idea is that stars can only form out
of cold, dense gas. Once the first stars have formed, they ionize and
heat the surrounding medium, so that no other, secondary stars can
form for a substantial amount of time. Put differently: wherever the
ionized bubbles extend, further star formation is
suppressed. Visualizing the network of radiation bubbles is therefore
crucial to characterize the strength of this negative feedback. More
precisely, there are two different kinds of radiation bubbles
surrounding the first stars: hard UV photons that are capable of
ionizing hydrogen atoms, and somewhat less energetic, soft UV photons
that are capable of dissociating hydrogen molecules. The latter are
very important, because they cool the primordial gas, thus enabling
the formation of the first stars. The extent of the soft-UV bubbles
thus delineates regions where no stars can form. We find the
surprising result that soft UV photons seem much less effective in
shutting off secondary star formation than was previously thought. Our
visualizations clearly show the relation of the two classes of UV
bubbles: The soft UV bubbles are larger than the hard UV ones, but,
surprisingly, only slightly so, as can be seen in
Figure~\ref{fig:h2frac}.  As this figure demonstrates, the smoothing
filter and transparency used in our visualization is essential for
drawing new cosmological insights.

A related question concerns the overall time evolution of early
re-ionization.  Traditionally, it has been argued that re-ionization
is a strictly monotonous process. The fraction of the volume filled by
ionized bubbles was thought to only increase with time. Our work has
shown that the early stages of re-ionization occur in a much more
complex fashion, where bubbles grow, disappear again, and are
eventually replaced by new bubbles. The time evolution, therefore, is
highly intermittent, and not at all monotonic. This effect can be seen
in Figure~\ref{fig:density}, where the ionized region of the first
star has been largely replaced by the ionized regions of the second
and third star.  Note that these observations cannot be made without
seeing the spatial relationships that are made clear in our
visualizations.

Our visualizations also have allowed us to understand this
intermittency more fully. Specifically, we find that the bubbles do
not disappear completely after the star that produced them has died;
instead, a remnant of reduced, but still significant, ionization
lingers on for a long time (see Figure~\ref{fig:three_cores}), so that
a given point in space can be influenced simultaneously by such a
``ghost'' bubble, together with a freshly formed bubble around a star
that is still alive, as can be seen in
Figures~\ref{fig:h2frac}~and~\ref{fig:density}. Again, the
observational signature form the early re-ionization, which depends on
the precise degree of ionization, looks substantially different
because of this ``ghost'' effect.  Transparency and smoothing are
necessary to observe these structural details.

An effect related to the intermittecy of the ionization at the early
stages of star formation is the production of a high molecular
hydrogen (H$_2$) fraction within the ``ghost'' ionized regions.  This
occurs because the small degree of ionization that lingers in these
regions allows for free electrons to catalyze the production of H$_2$.
The high fraction of H$_2$ that develops in one of these regions can
be seen in Figure~\ref{fig:h2frac}, where the highest H$_2$ fraction
is clearly shown within the ``ghost'' ionized region in the middle of
our simulation box.  Because molecules are effective coolants of the
gas, these regions may be the sites of continued star formation, a
further illustration of how the UV radiation from the first stars is
surprisingly ineffectual at suppressing star formation.

\section{Related Work}
\label{sec:related_work}

In this section, we discuss related cosmological visualizations, 
isosurface extraction techniques and interpolation techniques.
We motivate the choices we made when designing our technique.

\subsection{Cosmological Visualizations}

Researchers have often visualized cosmological phenomena, but they
usually visualize their particle simulation directly.  To our
knowledge, there is no work that summarizes the various techniques and
their applications.  We discuss the work most relevant to ours below.

ParaView~\cite{paraview} has been used recently to compare the results
of several cosmological simulations~\cite{ahrens:2006, heitmann:2007}.
These direct particle visualizations demonstrate the power and
flexibility of using a well-tested, open-source visualization package.
Indeed, the large community of developers that contribute to
open-source, general visualization packages help ensure that recent
visualization advances are included quickly.  Dedicated astronomical
or cosmological visualization packages often lack features or
robustness.  For instance, Partiview~\cite{partiview} lacks
isosurfacing capability and our data set was too large for
AstroMD~\cite{ferro:2004}.  AstroMD, like ParaView, is based on
VTK~\cite{vtk}, but its development has stagnated (last update in
2004) and it lacks the abilities and easy use of a more mature system.

\subsection{Isosurface Extraction}
\label{sec:iso_extraction}

Most isosurface extraction methods operate only on structured data
usually a structured or unstructured grid~\cite{livnat:2005}.  
Livnat~\cite{livnat:2005} and Sutton {\it et al}.~\cite{sutton:2000}
provide overviews of popular isosurface extraction techniques.

Marching Cubes~\cite{Lorensen:1987:MCA} is the most well-known
isosurface extraction techniques.  Marching Cubes is relatively simple
to implement, but it requires a structured grid from which to create
the isosurfaces.  The original implementation relies on a uniform
grid, but octree-based optimizations also
exist~\cite{Wilhelms:1992:OFF}.

Value-space decomposition techniques, such as
NOISE~\cite{Livnat:1996:ANO} and interval
trees~\cite{Cignoni:1996:OIE, Cignoni:1997:SUI}, can extract
isosurfaces from datasets that lack structure, as can the various
techniques of Co {\it et al}.~\cite{co:2004, co:2005} and Rosenthal
{\it et al}.~\cite{rosenthal:2006}.  Unfortunately, implementations of
these techniques are usually not freely available.

Point-based isosurfacing techniques exist~\cite{co:2003, livnat:2004,
vonrymon:2004}, but these techniques use ``point-based'' to refer to
their use of points rather than triangles as rendering primitives.
Our use of ``points'' refers to using the location of the particles
from the cosmological simulation as the location for rendering
primitives, regardless of what primitive is used.  We mention this to
avoid confusion over the overloaded ``points'' term.

\subsection{Interpolation Techniques}

Many interpolation techniques exist, and they can be organized by
complexity, accuracy and efficiency.  Franke~\cite{franke:1982}
summarizes several classes of interpolation techniques on scattered
data. Since our observations can tolerate a reasonable amount of
approximation, we select our interpolation based on computational
efficiency.

Interpolation techniques on scattered data can be divided into {\it
global} techniques that use all data points at each interpolation
point, and {\it local} techniques that limit the amount of data used
at each interpolation point~\cite{franke:1982}.  Our technique is
local, though it can be made global with the proper settings, as
discussed in Section~\ref{sec:interpolation}.

The interpolation we use~\cite{Jensen:1996:GIU, Jensen:1998:ESO} can
be characterized as a $n^{th}$-nearest neighbor density
estimate~\cite{fukunaga:1973} and as an localized inverse weighted
distance interpolation, or localized Shepard's
method~\cite{shepard:1968}.  Shepard's method has been the subject of
numerous analyses and extensions, for example~\cite{barnhill:1977,
barnhill:1980, gordon:1978, schumaker:1976}.

Of other interpolation methods, perhaps the most popular are basis
function methods, such as Hardy's multiquadric interpolation
method~\cite{hardy:1971}.  This interpolation can provide greater
accuracy and detail than inverse weighted distance
methods~\cite{franke:1982} and has been shown to always be
solvable~\cite{micchelli:1986}, but it carries higher complexity and
computational costs than localized methods.  This interpolation
has been expanded to cover other radial basis
functions~\cite{dyn:1986, Renka:1988:MIL} and
B-splines~\cite{lee:1997}.

\section{Conclusion and Future Work}
\label{sec:conclusion}

In this paper, we describe our method of visualizing cosmological
point-based datasets.  This method is simple, flexible, and leverages
existing visualization software.  We demonstrate that our
isosurface-based visualization provides detail and clarity superior to
that of a direct visualization of the particle data.

With this visualization we have gained a clearer view of the effects
of the radiation from the first stars, allowing a detailed description
of the early development of the ionized regions that are created
around these first sources of light, as well as illustrating how star
formation can continue despite the suppressive effects of the
radiation.  The details of early star formation that we are able to
glean using our method of visualization will allow us to make improved
predictions for what the {\it James Webb Space Telescope}, the key
astronomy mission of the next decade, will discover in the early
universe, probing the formative first billion years of cosmic history.

Further, we are applying our technique to the visualization of other
early universe phenomena.  We have begun to visualize a simulation of
the first supernova explosion~\cite{greif:2007}, and our technique has
produced similar improvement in our image quality and the insights
possible because of them.

\acknowledgements{ Thanks to Kelly Gaither, Greg S. Johnson and Romy
Schneider at the Texas Advanced Computing Center, and to the anonymous
reviewers for their helpful comments that improved this paper.  Our
cosmological particle simulation was funded in part by NASA {\it
Swift} grant NNG05GH54G.  }


\end{document}